	\documentstyle[prd,aps,preprint]{revtex}
\begin{document}
\title{
               Vacuum Oscillations and Future Solar Neutrino Experiments
}
\author{
               Naoya Hata                               \\
{\it
               Department of Physics,
               University of Pennsylvania               \\
               Philadelphia, Pennsylvania 19104--6396
}%
}
\date{
               February 17, 1994,  UPR-0605T
}

\maketitle

\begin{abstract}

Vacuum oscillations are considered for the combined solar neutrino
observations, including the Kamiokande II spectrum data and
incorporating theoretical uncertainties and their correlations.
Despite the conceptual difficulty of the fine tuning between the
neutrino parameters and the Sun-Earth distance, 2-flavor vacuum
oscillations provide phenomenologically acceptable solutions.  There
are allowed regions at 99\% C.L. for $\Delta m^2 = (0.45 - 1.2) \times
10^{-10} \; \mbox{eV}\,^2$ and $\sin^22\theta = 0.6 - 1$; the best fit
solution is $\chi^2 / \mbox{d.f.} = 19.2 / 16$, which is acceptable at
16\% C.L.  Oscillations for sterile neutrinos are, however,
excluded by the averaged data at 99.4\% C.L.  The vacuum oscillation
solutions predict characteristic energy spectrum distortions and
seasonal variations in Sudbury Neutrino Observatory, Super-Kamiokande,
and BOREXINO.  Those predictions are given in detail, emphasizing that
the vacuum solutions are distinguishable from the MSW solutions.

\end{abstract}
\pacs{PACS numbers: 96.60.Kx 12.15.Ff 14.60.Lm 14.60.Pq}           

\newpage

\section{Introduction}

The solar neutrino experiments of Homestake
\cite{Homestake,Homestake-update}, Kamiokande
\cite{Kamiokande-II,Kamiokande-III}, and the gallium experiments of
SAGE \cite{SAGE} and GALLEX \cite{GALLEX} show significant deficits of
the solar neutrino flux when compared to the standard solar model
predictions \cite{Bahcall-Pinsonneault,Turck-Chieze-Lopes}, as
summarized in Table~\ref{tab_solarnu-data}.  Among many theoretical
propositions to resolve the discrepancy, the astrophysical solutions
in general are strongly disfavored by the observations
\cite{Bethe-Bahcall,BKL,BHKL,Castellani-DeglInnocenti-Fiorentini,HBL}.
Assuming the standard properties of neutrinos, the SSMs are excluded
\cite{Bethe-Bahcall}.  Moreover a model independent analysis concluded
that the lower observed Homestake rate relative to the Kamiokande rate
is essentially incompatible with all astrophysical solutions
\cite{HBL}.  On the other hand, the neutrino oscillation hypothesis
--- the simplest particle physics solution requiring the minimal
extension of the standard model of particle physics --- fits all
observations; the Mikheyev-Smirnov-Wolfenstein (MSW) \cite{MSW}
solutions with $\Delta m^2 \sim 10^{-5} \; \mbox{eV}\,^2$ gives a
perfect description of the data, and the obtained mass range is
consistent with the general expectation of the grand unified theories
with the seesaw mechanism (for recent MSW analyses, see
Ref.~\cite{HL-analysis} and references therein).  Vacuum oscillations
\cite{vacuum-oscillations} also offer phenomenologically acceptable
solutions.  These, however, require a fine tuning between the
neutrino parameters and the Sun-Earth distance and suffer a conceptual
drawback.

In this paper the combined solar neutrino observations including the
Kamiokande II spectrum data are investigated for vacuum oscillations
(for other recent analysis, see
Ref.~\cite{Barger-Phillips-Whisnant,Acker-Pakvasa-Pantaleone,Krastev-Petcov,%
Acker-Balantekin-Loreti,Krastev-Petcov-93,%
Phillips}).
\footnote{
Vacuum oscillations were examined recently by Krastev and Petcov for
the averaged data of the experiments including the theoretical
uncertainties, and for the time-divided data without the theoretical
uncertainties \cite{Krastev-Petcov-93}.  Here vacuum
oscillations are studied without the time-divided data, but including
the Kamiokande II spectrum data and incorporating the theoretical
uncertainties, and the consequences for the future experiments are
discussed. }
Up-to-date experimental data are used.  The theoretical uncertainties
and their correlations are fully incorporated.  It is stressed that,
since some of the flux uncertainties are comparable to the
experimental uncertainties, and also they are strongly correlated
experiment to experiment and flux to flux, a careful treatment of the
theoretical uncertainties is necessary.  The omission of the
theoretical uncertainties and their correlations can lead to a
nontrivial change of the allowed parameter space
\cite{HL-analysis}.  From the parameter space obtained from the
existing experiments, one can give robust predictions for the next
generation experiments of Sudbury Neutrino Observatory,
Super-Kamiokande, BOREXINO, and ICARUS, which will start operating in
the mid and late 1990's.  In particular the global solutions predict
characteristic spectrum distortions in SNO and Super-Kamiokande and
drastic seasonal variations in BOREXINO.  Those
predictions are explicitly given, and it is emphasized that those
measurements distinguish the vacuum oscillation solutions from the MSW
solutions.

\section{Global Analysis for Vacuum Oscillations}

In 2-state oscillations in vacuum, the survival probability of the
electron neutrinos produced in the Sun and measured in a detector is
given by
\begin{equation}
     P_{\nu_e \rightarrow \nu_e}(E, r, R) =
                           1 - \sin^22\theta \sin^2 [(R-r) \Delta m^2 / 4E],
\label{eqn_P}
\end{equation}
where $\Delta m^2$ and $\theta$ are the squared mass difference and
the mixing angle between electron neutrinos and the other species of
neutrinos to which $\nu_e$'s oscillate; $E$ is the neutrino energy and
$r$ and $R$ are the neutrino production location and the detector
location measured from the center of the Sun.  As the reference SSM,
the latest model including the helium diffusion effect by Bahcall and
Pinsonneault is used \cite{Bahcall-Pinsonneault}.  In constraining the
parameter space from the observations, the survival probability
$P_{\nu_e \rightarrow \nu_e}(E, r, R)$ is convoluted with the neutrino
production profile functions \cite{Bahcall-Pinsonneault} and
integrated over $r$, and is also averaged over the Sun-Earth distance
$R$ when compared to the time-averaged rate of the experiments.  For
the continuous spectrum flux components the survival probability is
also integrated over $E$.  For the radiochemical detectors the
detector cross sections are taken from
Ref.~\cite{Bahcall-Ulrich,Bahcall-book} and included; for Kamiokande
the detector cross section, the detector resolution, and the detector
efficiency are included \cite{Kamiokande-II}.  When the Kamiokande II
spectrum data are incorporated, the electron spectrum are divided into
14 bins according to the published data \cite{Kamiokande-II}.

Since the theoretical uncertainty of the $^8$B flux (14\%) in the
Bahcall-Pinsonneault SSM is comparable to the experimental
uncertainties of Homestake (11\%) and Kamiokande (14\%), it is
important to incorporate the theoretical uncertainties of the neutrino
fluxes in the analysis.  The flux uncertainties are strongly
correlated experiment to experiment, and also flux to flux.  The SSM
uncertainties are parametrized with the uncertainties of the central
temperature and the uncertainties of the relevant nuclear reaction
cross sections, following the prescription described in
Ref.~\cite{HL-analysis}.  The uncertainties from the detector cross
sections are also included and are correlated between the two gallium
experiments.

First the analysis has been carried out for Homestake, SAGE, GALLEX,
and the averaged rate of Kamiokande II and III without the spectrum
data.  The two gallium experiments are treated as independent data
points.  When we confront observations to a theory, there are two
statistical questions to ask:
\begin{enumerate}

\item    Given the observations, is the theory acceptable or not?

\item    Once we decide that the theory is acceptable, what is the
	 allowed parameter space in the theory at a certain
	 confidence level?

\end{enumerate}
In the present case, the first question is quantified by evaluating
the goodness-of-fit (GOF) for the $\chi^2$ minimum in the $\Delta m^2 -
\sin^22\theta$ space.  The global $\chi^2$ minimum when the data are
fit to the oscillation parameters is 5.0 for 2 degrees of freedom ( 2
d.f. = 4 experiments $-$ 2 parameters).  The probability of obtaining
the $\chi^2$ minimum equal to or larger than 5.0 for 2 d.f.  by chance
is 8\%.  That is, the best fit solution for vacuum oscillations is
allowed at the 8\% confidence level (C.L.), which is not an excellent fit,
but acceptable.  There are seven local $\chi^2$ minima in the region
$\Delta m^2 = (0.35 - 1.3)\times 10^{-10} \, \mbox{eV}\,^2$ and
$\sin^22\theta = 0.6 - 1.0$; their oscillation parameters, $\chi^2$,
and the GOF values are summarized in Table~\ref{tab_bestfit-nospec}.

The data have also been fit to $\Delta m^2$, $\sin^22\theta$, {\it
and} the mean Sun-Earth distance simultaneously.  For the best fit
solution (Solution D in Table~\ref{tab_bestfit-nospec}), the mean
Sun-Earth distance has to be fine-tuned within 5\% to describe the
existing observations at 90\% C.L.  Unless we invoke some kind of
anthropic principle, there is no apparent logical connections between
the oscillation parameters and the Sun-Earth distance, and this fine
tuning is considered as a conceptual drawback in vacuum oscillations.

Once vacuum oscillations are accepted as a phenomenologically viable
theory for the solar neutrino observations, the C.L. regions are
obtained by evaluating $\chi^2$ at each point in the $\Delta m^2 -
\sin^22\theta$ space.   The allowed regions are defined by
\begin{equation}
         \chi^2 (\sin^22\theta, \Delta m^2)
     \le \chi_{\mbox{\scriptsize min}} + \Delta \chi^2,
\end{equation}
where $\chi_{\mbox{\scriptsize min}}$ is the global $\chi^2$ minimum,
and $\Delta \chi^2 =$ 4.6, 6.0, and 9.2 for 90, 95, and 99\% C.L.,
respectively.
\footnote{%
This definition of the allowed parameter space assumes a gaussian
distribution of the probability density around the global $\chi^2$
minimum.  That is, the allowed regions should have elliptical shape,
and this is only a crude approximation in the present case.  Therefore
the allowed regions shown in Fig.~\ref{fig_fit-each},
\ref{fig_comb-nospec}, and \ref{fig_comb-spectrum} should be
taken as qualitative displays of confidence levels.} %
The constraints obtained from each Homestake, Kamiokande, and the
combined gallium experiment at 95\% C.L. are shown in
Fig,~\ref{fig_fit-each}.

When the parameters are fit with the combined observations, the
allowed regions are essentially controlled by the Homestake data and
limited to $\Delta m^2 = ( 0.35 - 1.3 ) \times 10^{-10} \, \mbox{eV}
\,^2$ and $\sin^22\theta = 0.6 - 1.0$.  The regions allowed by the
combined observations at 90, 95, and 99\% C.L. are shown in
Fig.~\ref{fig_comb-nospec}; the allowed regions are essentially in
agreement with the recent analysis by Krastev and Petcov
\cite{Krastev-Petcov-93}.  The slight differences are mainly due to
the difference in the input data.

Since the $\nu_e$ survival probability in this parameter space (i.e.,
the Sun-Earth distance $\sim$ the oscillation length) are often
energy dependent, the analysis has been also carried out with the
additional Kamiokande II energy spectrum data.  The 14 data points
obtained in the Kamiokande II experiment are used
\cite{Kamiokande-II}; the theoretical uncertainties as well as the
experimental systematic uncertainties (0.06 $\times$ SSM) are
correlated among the 14 bins.  No spectrum data are available from
Kamiokande III and the average value is used as an independent data
point.  The SSM flux uncertainties are also correlated between
Kamiokande II and III.
\footnote{
The possible correlations of the systematic uncertainties in Kamiokande
II and Kamiokande III were ignored.  }
The expected spectrum distortions for three local minima (B, D, and F)
for Kamiokande along with the actual Kamiokande II data are shown in
Fig.~\ref{fig_KamIIspec}.  The global $\chi^2$ minimum of the combined
analysis is 19.2 for 16 d.f. ( = 18 data points -- 2 parameters),
which is acceptable at 84\% C.L.  The neutrino parameters and the GOF
for all local $\chi^2$ minima in this region are listed in
Table~\ref{tab_bestfit}.  The allowed parameter regions are displayed
in Fig.~\ref{fig_comb-spectrum}.  The parameter space allowed at 99\%
C.L.  is for
\begin{equation}
     \Delta m^2 = ( 0.45 - 1.2 ) \times 10^{-10} \, \mbox{eV} \,^2
     \mbox{  and  }
     \sin^22\theta = 0.6 - 1.0,
\end{equation}
and remain essentially unchanged by including the Kamiokande II
spectrum information.

Vacuum oscillations can also be considered between electron neutrinos
and sterile neutrinos
\cite{sterile,Barger-Phillips-Whisnant,Krastev-Petcov,Krastev-Petcov-93}.
In this case there are no neutral current events from the transformed
neutrinos in Kamiokande, which would be present and contribute about
15\% of the total signal in the case of oscillations into $\nu_\mu$ or
$\nu_\tau$.  This absence of the neutral current contributions in
Kamiokande makes the difference between the Kamiokande rate and the
Homestake rate effectively wider, and the fits become considerably
worse.  In the joint $\chi^2$ analysis without the Kamiokande spectrum
data, the global $\chi^2$ minimum is 10.1 for 2 d.f., which is an
extremely poor fit.  The GOF is 0.63\%, and the solution is excluded
at 99.4\% C.L.  That is, including all theoretical and
experimental uncertainties, the existing experiments are essentially
incompatible with vacuum oscillation for sterile neutrinos.
\footnote{
The same conclusion was obtained by Krastev and Petcov
\cite{Krastev-Petcov-93}, but with less statistical significance.
The difference is entirely due to the difference in the input data.  }
When the Kamiokande II spectrum data are included, the GOF improves
(4.8\%), but this is an artifact of increasing the number of degrees
of freedom by adding the Kamiokande II 14 bins, and does not mean that
the sterile oscillations are favored by the Kamiokande spectrum data;
the incompatibility of the data and the sterile oscillations remains
essentially unchanged.  The oscillation parameters and GOF for each
$\chi^2$ local minimum is listed in
Table~\ref{tab_bestfit-nospec-sterile} for the averaged rate, and in
Table~\ref{tab_bestfit-sterile} for the analysis including the
Kamiokande II spectrum data.

\section{Predictions for the Future Solar Neutrino Experiments}

The oscillation hypothesis for the solar neutrinos, either the MSW
mechanism or the vacuum oscillations, will be verifiable in the
next-generation solar neutrino experiments of SNO \cite{SNO},
Super-Kamiokande \cite{Super-Kamiokande}, BOREXINO \cite{BOREXINO},
and ICARUS \cite{ICARUS}.  Moreover those experiments should be able
to distinguish the vacuum solutions and the MSW solutions, and
determine the oscillation parameters accurately.  The key measurements
in those experiments are:
\begin{enumerate}
\item  	The ratio of charged and neutral current events (CC/NC)
	from the $^8$B neutrinos in SNO.
\item  	The energy spectrum measurements in SNO and
	Super-Kamiokande.
\item   The measurements of time variations of the $^8$B
	neutrinos in SNO and Super-Kamiokande, and the measurements
        of time variations  of the $^7$Be neutrinos in BOREXINO.
\end{enumerate}

The CC/NC measurement in SNO is the gold plated observable in
confirming or falsifying the oscillation hypothesis.  The comparison
of the observed value (CC/NC)$_{\mbox{\scriptsize SNO}}$ and the SSM
value (CC/NC)$_{\mbox{\scriptsize SSM}}$ is free from the SSM flux
uncertainties, and a depletion of the observed charged current
signifies the oscillations between $\nu_e$ and $\nu_\mu$ or
$\nu_\tau$.  Assuming the Bahcall-Pinsonneault SSM, the currently
allowed parameter space at 95\% C.L. predicts
\begin{equation}
	\frac{ \mbox{(CC/NC)}_{\mbox{\scriptsize SNO}}}
	     { \mbox{(CC/NC)}_{\mbox{\scriptsize SSM}}}
	=   0.15 - 0.5.
\end{equation}
There are two caveats, however.  The CC/NC measurement by itself does
not distinguish the type of oscillations (e.g., whether the MSW effect
or vacuum oscillations).  Also (CC/NC)$_{\mbox{\scriptsize SNO}}$ is
the same as (CC/NC)$_{\mbox{\scriptsize SSM}}$ if the oscillations are
for sterile neutrinos.  The measurements of the energy spectrum and of
time variations are important in determining those detailed features
of neutrino oscillations.  In the following, the predictions for
spectrum distortions and the time variations are discussed separately,
but those two effects are often strongly correlated, and the real data
in the future experiments should be analyzed accordingly.

Although the existing Kamiokande energy spectrum data do not
significantly constrain the parameter space, the next generation
experiments with high statistics and with lower energy thresholds
($\sim$ 5 MeV) are quite sensitive to the spectrum distortions caused
by oscillations.  In Fig.~\ref{fig_sno-spectrum}, the expected
spectrum shapes are displayed for three vacuum solutions (B, D, and F
defined in Table~\ref{tab_bestfit}) along with the spectra expected
from the MSW solutions obtained by the global analysis; the MSW
large-angle solution has little distortion and is identical to the
astrophysical solutions.  The spectra are normalized above the
proposed detector threshold (5 MeV).  The charged current cross
section \cite{Nozawa,nu-d-cross-section} and the detector resolutions
\cite{SNO} are also included.
\footnote{%
In the SNO spectrum analysis, it is important to include the effect of
the continuous spectrum of the out-going electrons and also the
detector resolution: the approximation of the out-going electron
spectra from the monoenergetic neutrinos with a delta function (i.e.,
``observed electron energy'' = ``neutrino energy'' -- 1.44 MeV)
significantly overestimates the detector sensitivity for spectrum
distortions and is inappropriate for the spectrum shape analysis.  } %
The error bars correspond to the statistical uncertainties for a total
of 6,000 events (equivalent to 2 year operation).  The vacuum
oscillations are clearly distinguishable from the MSW solutions and
the astrophysical solutions.  The similar predictions for
Super-Kamiokande are displayed in Fig.~\ref{fig_superk-spectrum} with
error bars corresponding to a total of 16,000 events (2 year
operation).

The survival probability [Eqn.~(\ref{eqn_P})] is also sensitive to the
Sun-Earth distance $R$ and the solutions of the combined observation
predict seasonal variations of the solar neutrino signal due to the
eccentricity of the Earth orbit.  Especially drastic time variations
are expected for the $^7$Be neutrino signals since they are
mono-energetic (therefore not smeared out by the energy integral), and
also the oscillation length is shorter than for most of the $^8$B
neutrinos \cite{Barger-Phillips-Whisnant,Krastev-Petcov,%
Krastev-Petcov-93}.  For the existing experiments the expected time
variations for the global solutions (B, D, and F) are displayed in
Fig.~\ref{fig_time-variation-now}: the error bars are the statistical
uncertainties when (overly optimistic) large number of events are
assumed: 500, 1,000, and 1,000 events for Kamiokande, Homestake, and
the gallium experiment, while the currently accumulated signals are
$\sim$ 200, $\sim500$, and $\sim$ 50 events, respectively.  Obviously
there is little chance of detecting the seasonal variations by the
currently operating experiments.

The time variations expected in SNO, Super-Kamiokande, and BOREXINO
are shown in Fig.~\ref{fig_time-variation-future}: the error bars are
for the total number of events 12,000, 32,000, and 12,000,
respectively (each corresponds to 4 years of operation).  The seasonal
variations should be observable in these experiments.  In particular
the time variations are most drastic in BOREXINO because of its
sensitivity for the $^7$Be neutrinos.  Those measurements are
unmistakable and should provide strong constraints on the oscillation
parameters.

\section{Conclusion}

The Homestake, SAGE, GALLEX and Kamiokande data including the
Kamiokande II spectrum data are analyzed for neutrino oscillations in
vacuum.  There are phenomenologically acceptable solutions in $\Delta
m^2 = (0.45 - 1.2) \times 10^{-10} \, \mbox{eV}\,^2 $ and
$\sin^22\theta = 0.6 - 1.0$, although these solutions suffer from the
fine tuning between the oscillation parameters and the Sun-Earth
orbit.  Oscillations for sterile neutrinos are strongly disfavored by
the observations and are excluded at 99.4\% C.L.  The acceptable
solutions for the flavor oscillations predict characteristic energy
spectrum distortions and seasonal variations in the next generation
solar neutrino experiments.  The measurements of these signals should
distinguish the vacuum oscillations from the astrophysical solutions
as well as from the MSW solutions.

\acknowledgements

I thank Paul Langacker for his encouragement, useful comments, and
careful reading of the manuscript.  It is also pleasure to thank
Eugene Beier and Sidney Bludman for useful discussions.  I thank
Plamen Krastev for directing my attention to the importance of
measuring the time variations of the $^7$Be flux.  This work was
supported by the Department of Energy Contract No.\
DE-AC02-76-ERO-3071.




\begin{table}[p]
\caption{
%
%
The standard solar model predictions of Bahcall and Pinsonneault
\protect\cite{Bahcall-Pinsonneault} (BP SSM) and of Turck-Chi\`eze
and Lopes \protect\cite{Turck-Chieze-Lopes} (TCL SSM).  The
Bahcall-Pinsonneault model includes the helium diffusion effect, while
the Turck-Chi\`eze--Lopes model does not.  Also shown are the results
of the solar neutrino experiments.  The gallium experiment is the
combined result of SAGE and GALLEX I and II.
}
\label{tab_solarnu-data}
\vspace{1.0ex}
\begin{tabular}{ l  c c c }
%
               & BP SSM        & TCL SSM      & Experiments \\
\hline
Kamiokande \tablenotemark[1]
               &  1 $\pm$ 0.14 & 0.77 $\pm$ 0.19 & 0.51 $\pm$ 0.07 BP SSM  \\
Homestake \tablenotemark[2] (SNU)
               &  $8\pm 1$   & 6.4 $\pm$ 1.4  & 2.34 $\pm$ 0.26
                                                (0.29 $\pm$ 0.03 BP SSM) \\
Ga experiment (SNU)
               & 131.5 $^{+7}_{-6}$ & 122.5 $\pm$ 7 & 81 $\pm$ 13
                                                    (0.62 $\pm$ 0.10 BP SSM) \\
SAGE \tablenotemark[3] (SNU)    & &  & 70 $\pm$ 22 (0.53 $\pm$ 0.16 BP SSM) \\
GALLEX \tablenotemark[4] (SNU)  & &  & 87 $\pm$ 16 (0.66 $\pm$ 0.12 BP SSM)
\end{tabular}
\tablenotetext[1]{%
The result of the combined data of 1040 days of Kamiokande II
[0.47 $\pm$ 0.05 (stat) $\pm$ 0.06 (sys) BP SSM]
\cite{Kamiokande-II}
and 514.5 days of Kamiokande III
[0.57 $\pm$ 0.06 (stat) $\pm$ 0.06 (sys) BP SSM]
\cite{Kamiokande-III}.
}
\tablenotetext[2]{%
The result of Run 18 to 122 is
2.34 $\pm$ 0.16 (stat) $\pm$ 0.21 (sys) SNU  \cite{Homestake-update}.
}
\tablenotetext[3]{%
The preliminary result of SAGE I (from January, 1990 through May, 1992) is
70 $\pm$ 19 (stat) $\pm$ 10 (sys) SNU \cite{SAGE}.
}
\tablenotetext[4]{%
The combined result of GALLEX I and II (including 21 runs through April, 1993)
is 87 $\pm$ 14 (stat) $\pm$ 7 (sys) SNU \cite{GALLEX}.
}

\end{table}

\begin{table}[p]
\centering
\caption{
%
%
The solutions for vacuum oscillations when the combined result of
Homestake, Kamiokande, SAGE, and GALLEX is used.  The SAGE and GALLEX
data are fit separately.  The Kamiokande II spectrum data are not
included.  A -- G correspond to seven local $\chi^2$ minima within the
allowed regions at 99\% C.L.  The goodness-of-fit (GOF) is the
probability of obtaining by chance a $\chi^2$ minimum value equal to or
larger than the value we actually obtained.}
\label{tab_bestfit-nospec}
\vspace{0.5ex}
\begin{tabular}{l  c c c c}
   &   $\Delta m^2$ (eV$^2$) & $\sin^22\theta$ & $\chi^2$ (2 d.f.) & GOF (\%)\\
\hline 
 A   &    $1.2 \times 10^{-10}$    &    1.00    &       10.9  &  0.42 \\
 B   &    $1.0 \times 10^{-10}$    &    0.96    &       5.7   &  5.9  \\
 C   &    $9.0 \times 10^{-11}$    &    0.84    &       5.1   &  7.9  \\
 D   &    $7.9 \times 10^{-11}$    &    0.78    &       5.0   &  8.0  \\
 E   &    $6.3 \times 10^{-11}$    &    0.81    &       5.2   &  7.6  \\
 F   &    $5.2 \times 10^{-11}$    &    0.99    &       7.5   &  2.3  \\
 G   &    $3.8 \times 10^{-11}$    &    1.00    &       14.5  &  0.071 \\
\end{tabular}
\end{table}

\vspace{10ex}
\begin{table}[p]
\centering
\caption{
%
%
Same as Table~\protect{\ref{tab_bestfit-nospec}} except that the
Kamiokande II spectrum data are included.  }
\label{tab_bestfit}
\vspace{0.5ex}
\begin{tabular}{l  c c c c}
   &   $\Delta m^2$ (eV$^2$) & $\sin^22\theta$ & $\chi^2$ (16 d.f.) &GOF (\%)\\
\hline 
 A   &    $1.2 \times 10^{-10}$    &    1.00    &       25.4  &  3.1 \\
 B   &    $1.0 \times 10^{-10}$    &    0.91    &       20.5  &  12  \\
 C   &    $9.1 \times 10^{-11}$    &    0.82    &       19.2  &  16  \\
 D   &    $7.8 \times 10^{-11}$    &    0.78    &       19.2  &  16  \\
 E   &    $6.5 \times 10^{-11}$    &    0.80    &       20.4  &  12  \\
 F   &    $5.2 \times 10^{-11}$    &    0.99    &       23.7  &  4.9  \\
 G   &    $3.8 \times 10^{-11}$    &    1.00    &       29.9  &  0.79 \\
\end{tabular}
\end{table}

\begin{table}[p]
\centering
\caption{
%
%
The solutions for vacuum oscillations for sterile neutrinos when the
combined result of Homestake, Kamiokande, SAGE, and GALLEX is used.
The SAGE and GALLEX data are fit separately.  The Kamiokande II
spectrum data are not included.  A -- G correspond to seven local
$\chi^2$ minima within the region.}
\label{tab_bestfit-nospec-sterile}
\vspace{0.5ex}
\begin{tabular}{l  c c c c}
   &   $\Delta m^2$ (eV$^2$) & $\sin^22\theta$ & $\chi^2$ (2 d.f.) & GOF (\%)\\
\hline 
 A   &    $1.2 \times 10^{-10}$    &    1.00    &       13.5  &  0.12 \\
 B   &    $1.1 \times 10^{-10}$    &    0.88    &       10.1  &  0.63  \\
 C   &    $9.1 \times 10^{-11}$    &    0.79    &       11.0  &  0.42  \\
 D   &    $7.8 \times 10^{-11}$    &    0.75    &       11.0  &  0.41  \\
 E   &    $6.3 \times 10^{-11}$    &    0.78    &       10.4  &  0.56 \\
 F   &    $5.0 \times 10^{-11}$    &    0.93    &       12.7  &  0.18  \\
 G   &    $3.8 \times 10^{-11}$    &    1.00    &       16.5  &  0.026 \\
\end{tabular}
\end{table}

\vspace{10ex}
\begin{table}[p]
\centering
\caption{
%
%
Same as Table~\protect{\ref{tab_bestfit-nospec-sterile}} except that
the Kamiokande II spectrum data are included.
}
\label{tab_bestfit-sterile}
\vspace{0.5ex}
\begin{tabular}{l  c c c c}
   &   $\Delta m^2$ (eV$^2$) & $\sin^22\theta$ & $\chi^2$ (16 d.f.) &GOF (\%)\\
\hline 
 A   &    $1.2 \times 10^{-10}$    &    0.98    &       30.1  &  1.7 \\
 B   &    $1.1 \times 10^{-10}$    &    0.84    &       27.0  &  4.2 \\
 C   &    $9.1 \times 10^{-11}$    &    0.78    &       27.6  &  3.5  \\
 D   &    $7.8 \times 10^{-11}$    &    0.74    &       27.1  &  4.1 \\
 E   &    $6.3 \times 10^{-11}$    &    0.77    &       26.4  &  4.8  \\
 F   &    $5.0 \times 10^{-11}$    &    0.91    &       29.3  &  2.2  \\
 G   &    $3.8 \times 10^{-11}$    &    1.00    &       32.3  &  0.92 \\
\end{tabular}
\end{table}


\begin{figure}[p]
\vspace{2ex}
\caption{
%
%
The allowed parameter space at 95\% C.L. for the Homestake, Kamiokande
and combined gallium experiments of SAGE and GALLEX including the
theoretical uncertainties and their correlations.  The Kamiokande II
spectrum data are not included.
}
\label{fig_fit-each}
\end{figure}

\begin{figure}[p]
\vspace{2ex}
\caption{
%
%
The parameter space allowed by the combined observation of the
Homestake, Kamiokande, and the combined gallium experiments of SAGE and
GALLEX, including the theoretical uncertainties and their
correlations.  The Kamiokande II spectrum data are not included.
}
\label{fig_comb-nospec}
\end{figure}

\begin{figure}[p]
\vspace{2ex}
\caption{
%
%
The spectrum shape divide by the SSM spectrum expected in Kamiokande.
The shape is shown for the combined solutions B, D, and F, which are
listed in Table~\protect\ref{tab_bestfit}.  Also shown are the Kamiokande
II spectrum data \protect\cite{Kamiokande-II}
}
\label{fig_KamIIspec}
\end{figure}

\begin{figure}[p]
\vspace{2ex}
\caption{
%
%
The parameter space allowed by the combined observation of the
Homestake, Kamiokande, and the combined gallium experiments of SAGE and
GALLEX, including the theoretical uncertainties and their
correlations.  The Kamiokande II spectrum data are included.
}
\label{fig_comb-spectrum}
\end{figure}

\begin{figure}[p]
\vspace{2ex}
\caption{
%
%
The electron spectrum shape expected in SNO charged current reaction
for the vacuum solutions B, D, and F (defined in
Table~\protect\ref{tab_bestfit}).  Also shown are the spectra expected
for the MSW solutions obtained from the combined observations.  The
spectrum shape of the astrophysical solutions (or no oscillations) is
identical to the MSW large-angle spectrum.  The spectra are normalized
above the threshold (5 MeV).  The error bars correspond to the
statistical uncertainties for a total of 6,000 events (2 year
operation).
}
\label{fig_sno-spectrum}
\end{figure}

\begin{figure}[p]
\vspace{2ex}
\caption{
%
%
The electron spectrum shape expected in Super-Kamiokande for the
vacuum solutions B, D, and F (defined in
Table~\protect\ref{tab_bestfit}).  Also shown are the spectra expected
for the MSW solutions obtained from the combined observations.  The
spectrum shape of the astrophysical solutions (or no oscillations) is
identical to the MSW large-angle spectrum.  The spectra are normalized
above the threshold (5 MeV).  The error bars correspond to the
statistical uncertainties for a total of 16,000 events (2 year
operation).
}
\label{fig_superk-spectrum}
\end{figure}

\begin{figure}[p]
\vspace{2ex}
\caption{
%
%
The seasonal variation of the signal expected in the existing
experiments for the vacuum solutions B, D, and F (see
Table~\protect\ref{tab_bestfit}).  The error bars corresponds to
(overly) optimistic statistical errors.  The accumulated events in the
existing experiments are $\sim$ 200, $\sim$ 500, and $\sim$ 50 for
Kamiokande, Homestake, and the combined gallium experiments.  No
significant constraint is expected from the existing experiments.
}
\label{fig_time-variation-now}
\end{figure}

\begin{figure}[p]
\vspace{2ex}
\caption{
%
%
The seasonal variation of the signal expected in the future
experiments of SNO (charged-current reaction), Super-Kamiokande, and
BOREXINO for the vacuum solutions B, D, and F (see
Table~\protect\ref{tab_bestfit}).  The error bars correspond to the
total events for 4 year operation.  Those measurements should be
sensitive enough to distinguish the vacuum oscillations from the
MSW solutions and constrain the parameter space.
}
\label{fig_time-variation-future}
\end{figure}

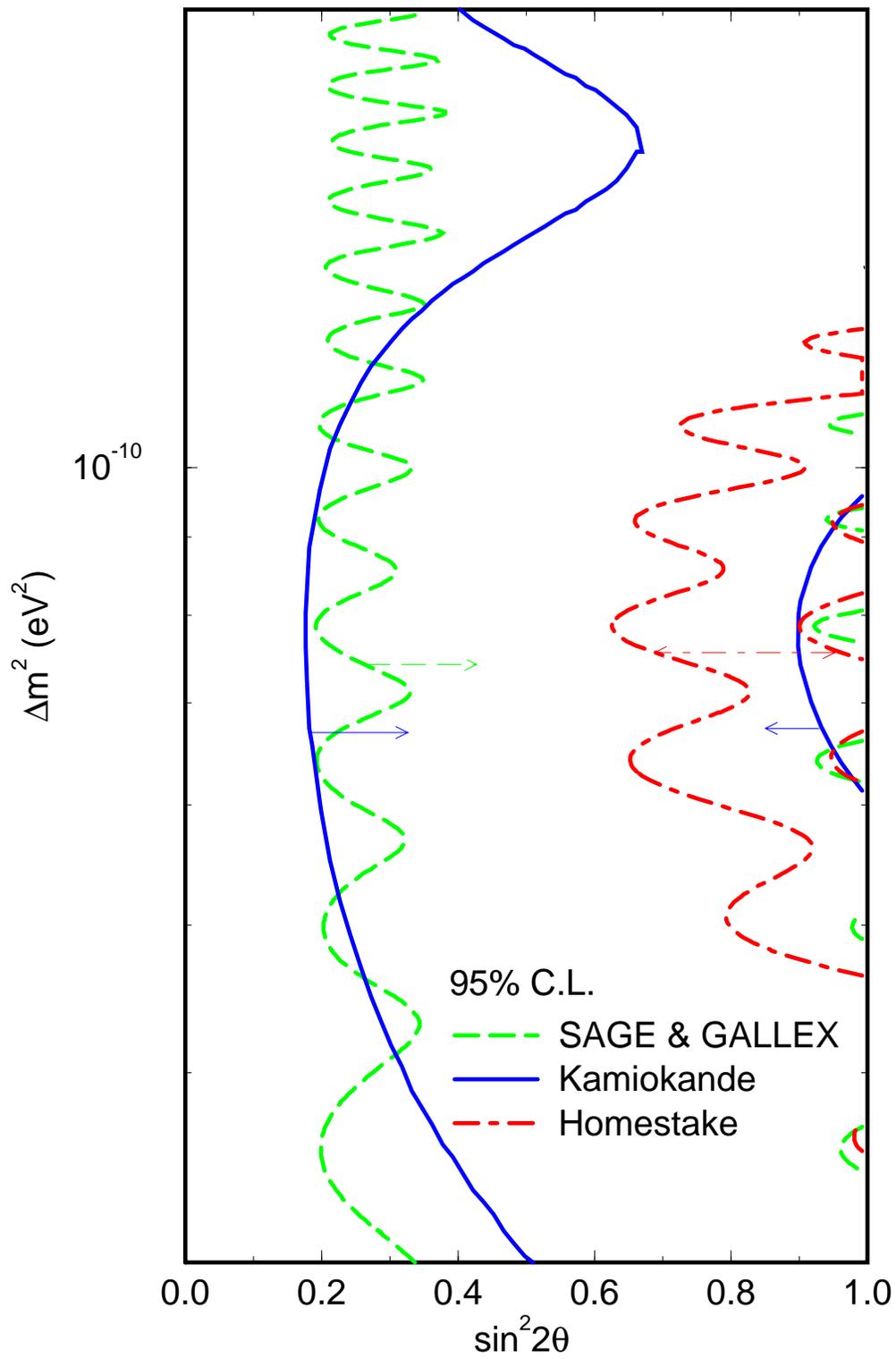
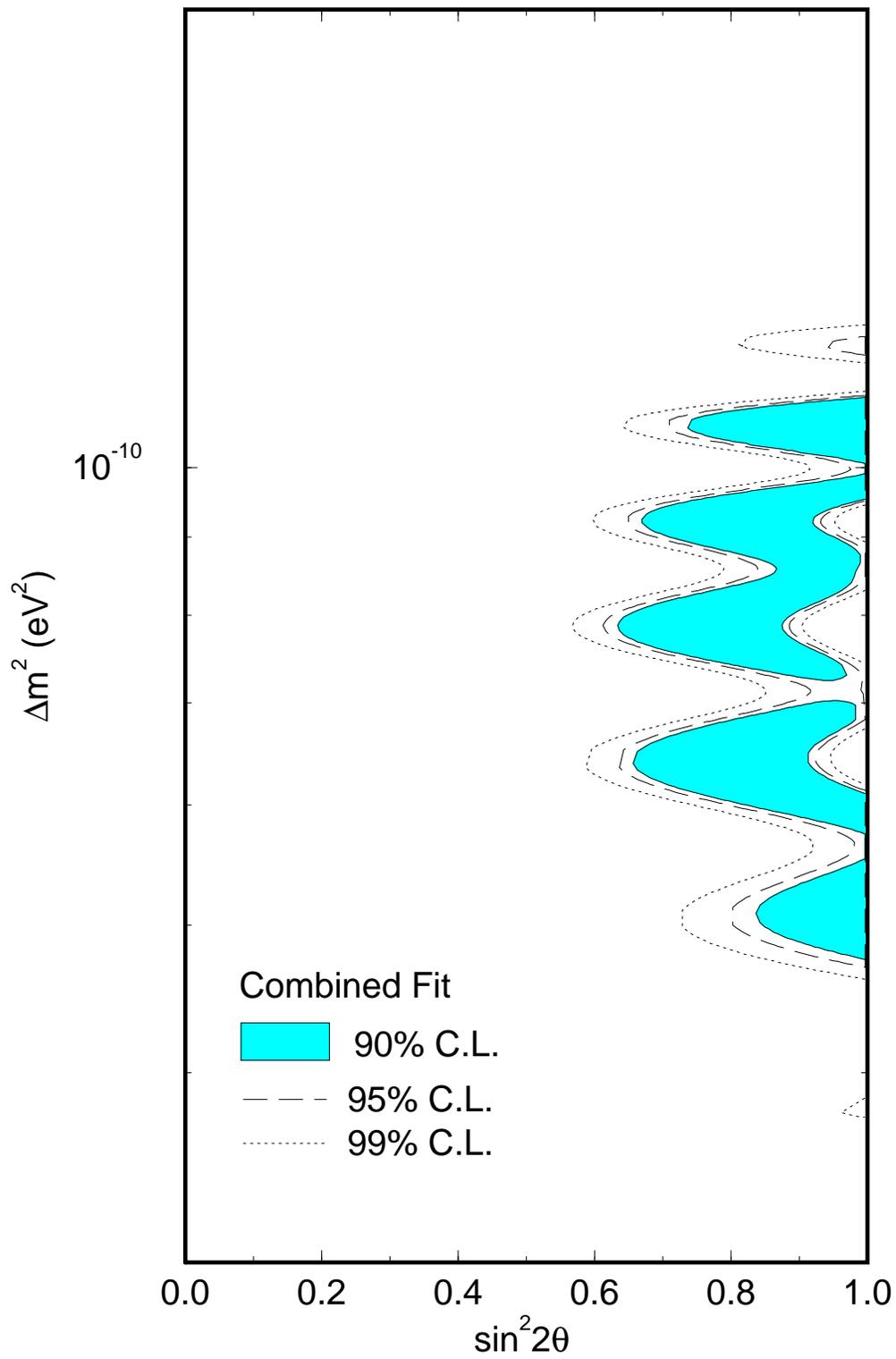
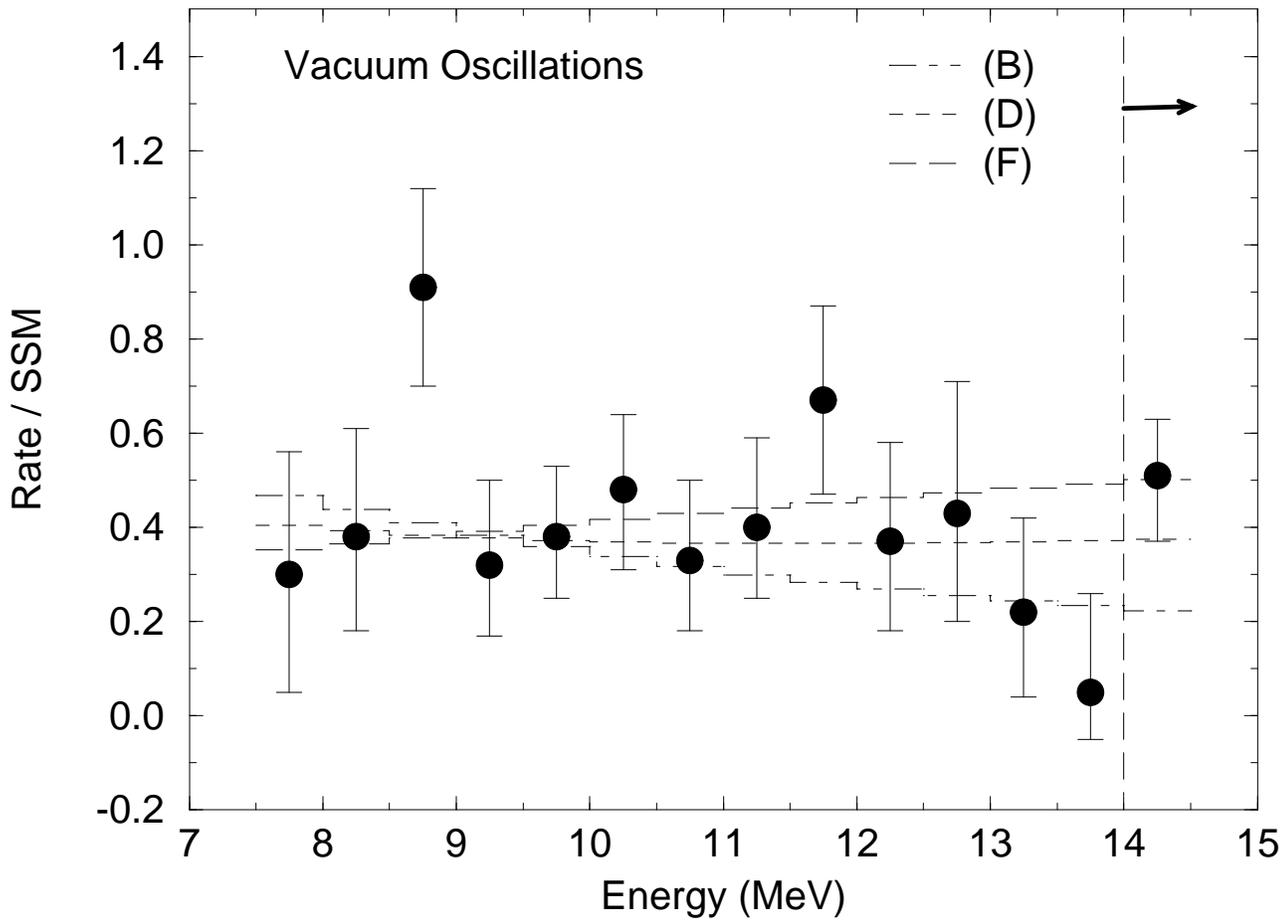
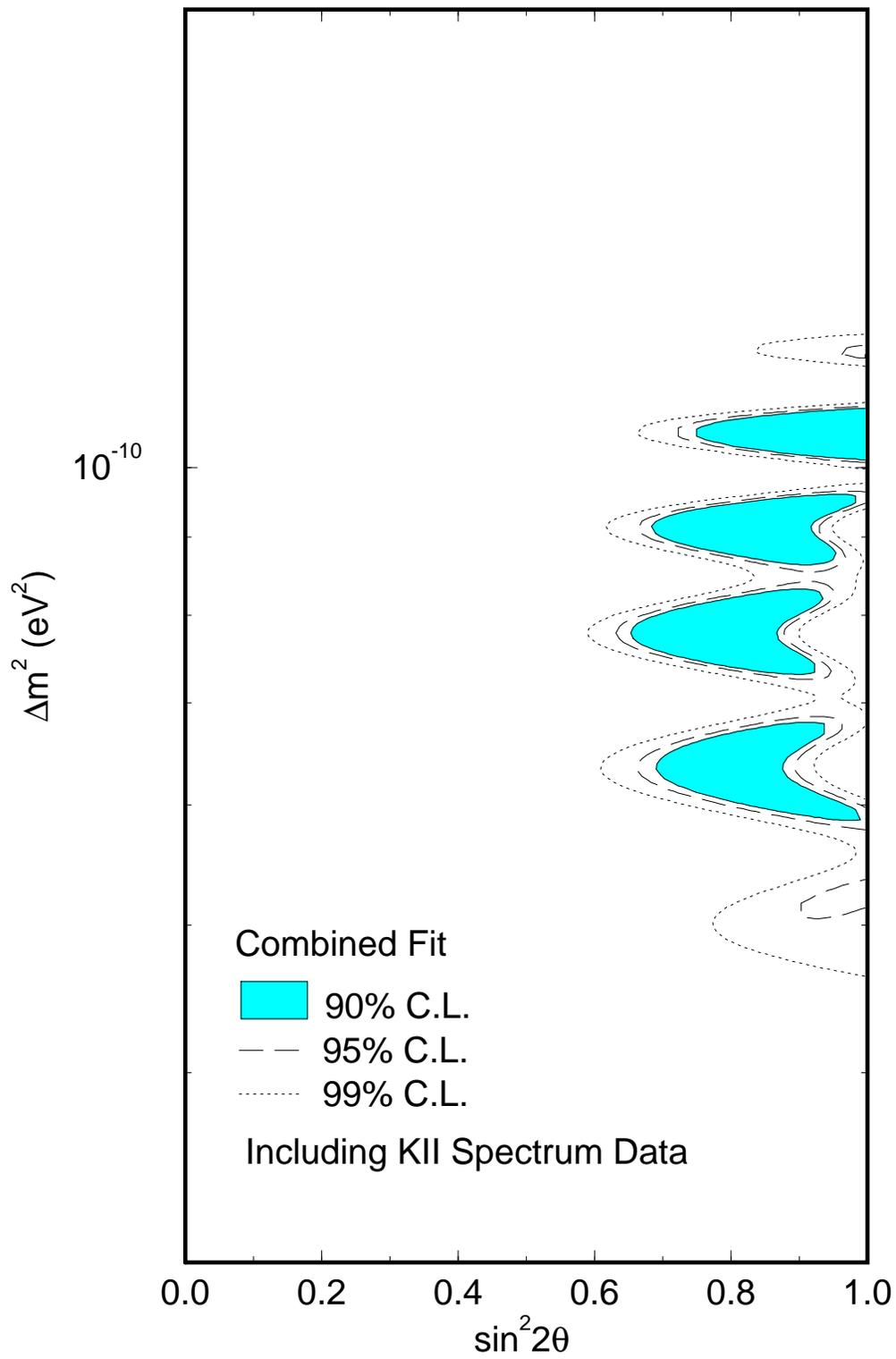
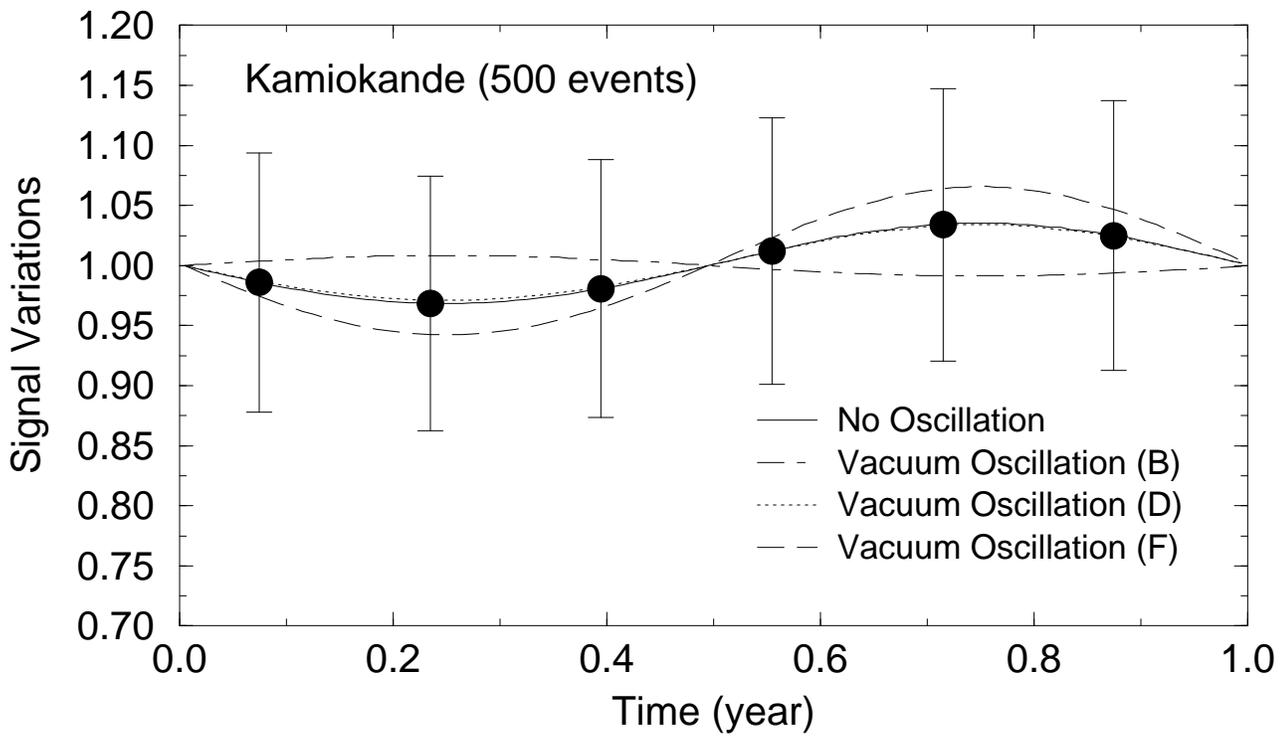
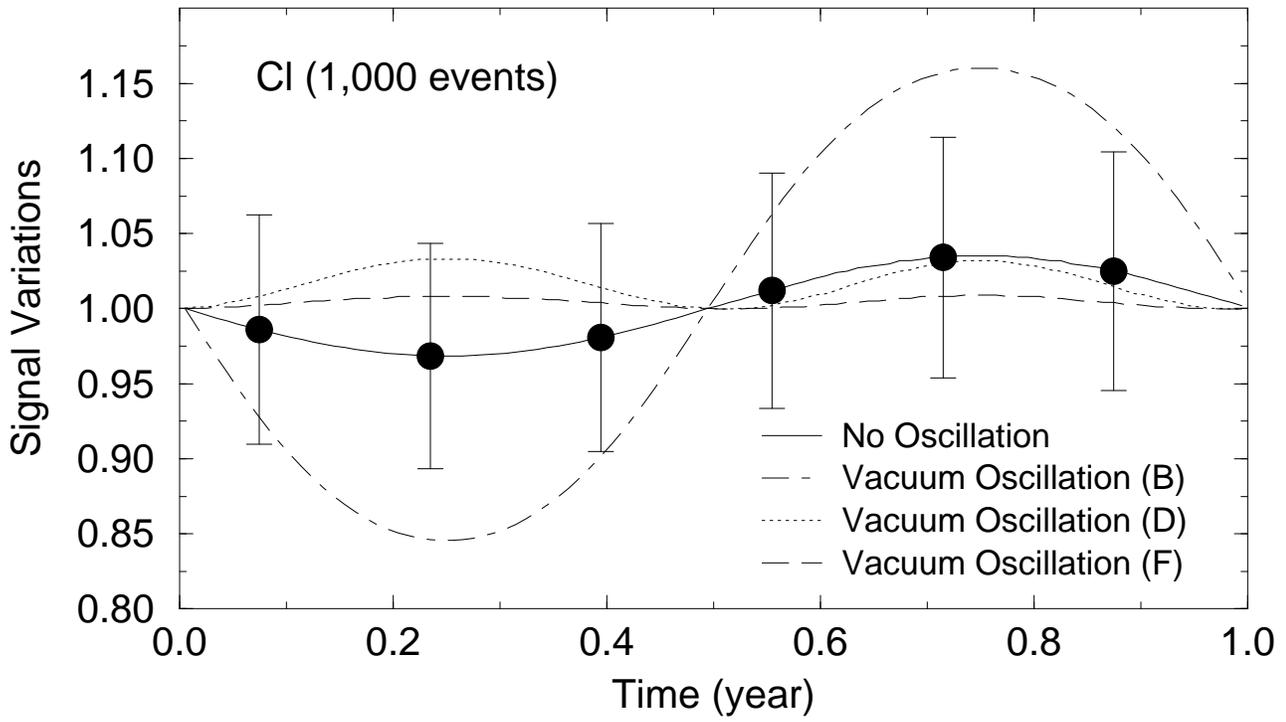
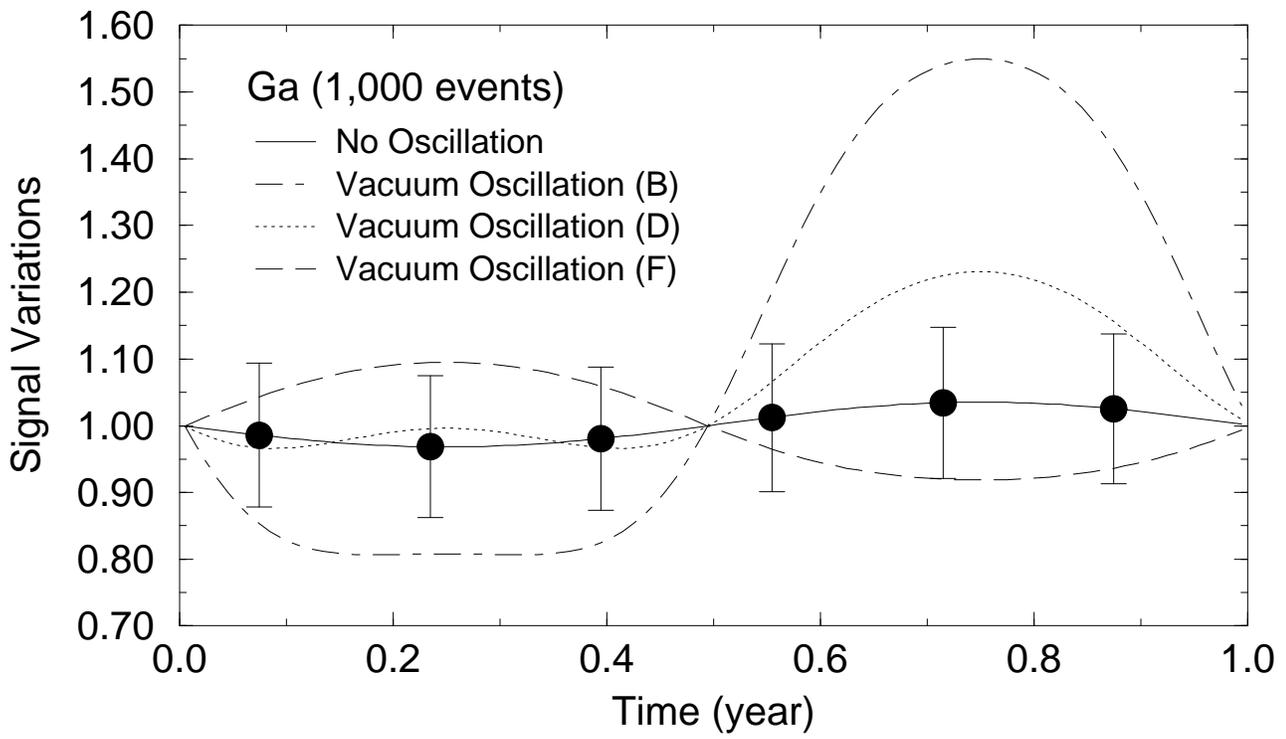
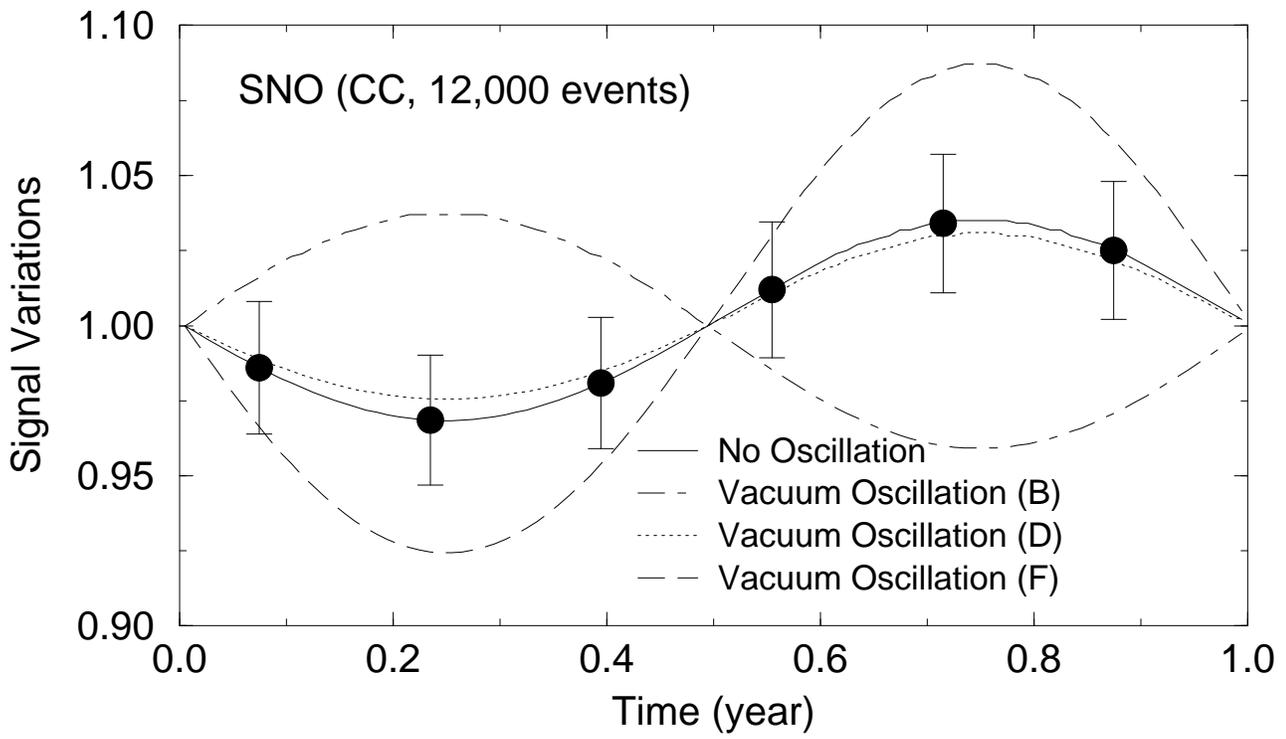
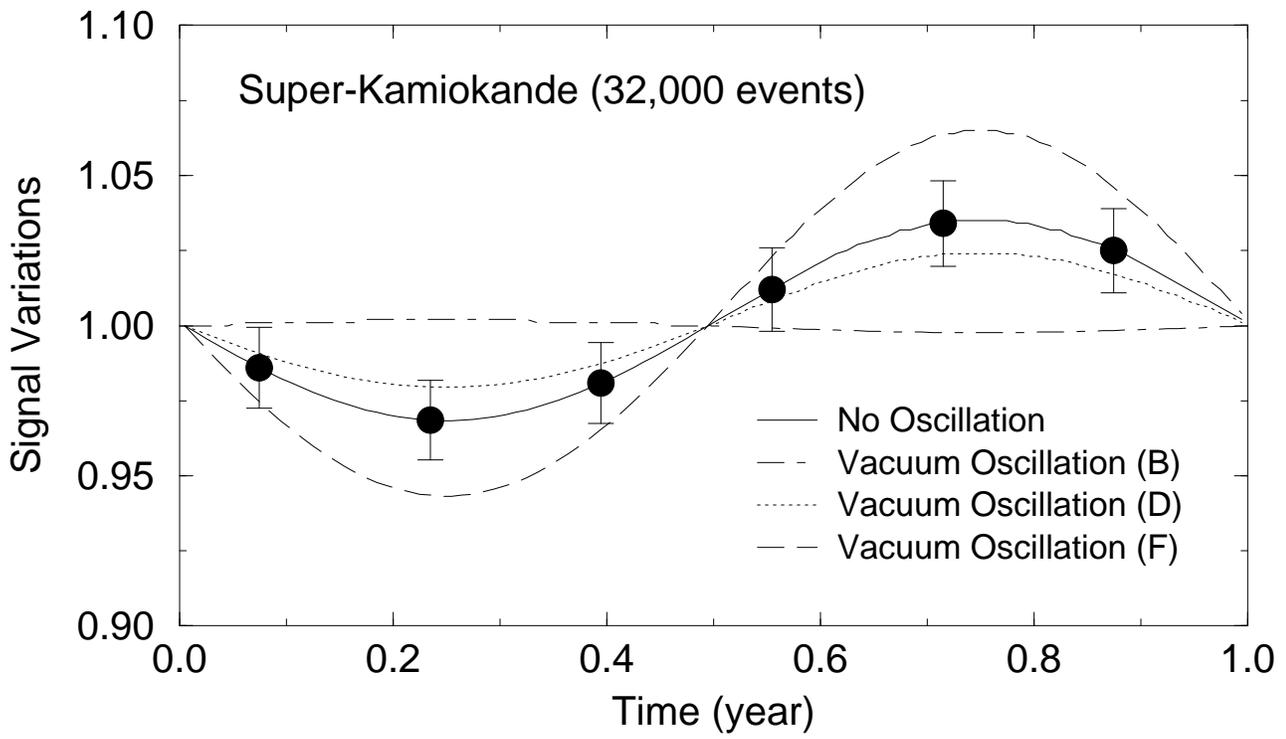
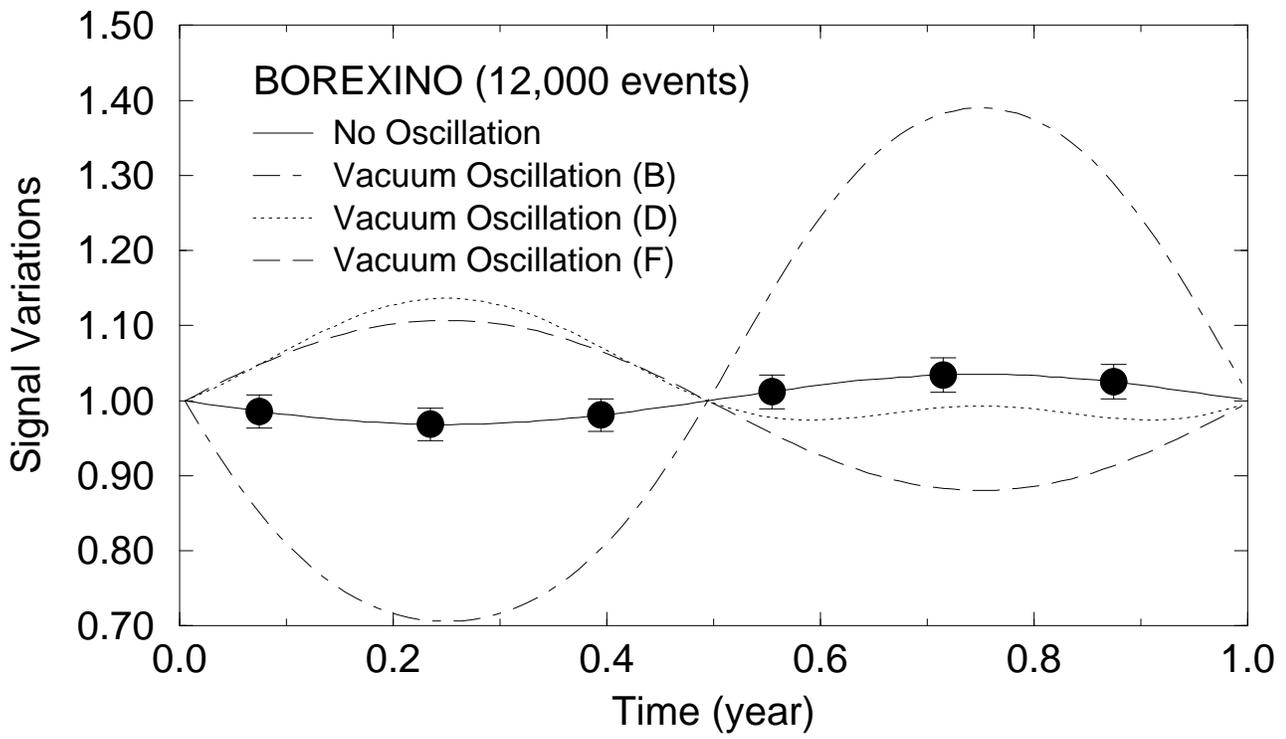
\end{document}